\documentclass[letter]{aa}  
\usepackage{graphicx}
\usepackage{textcomp}
\usepackage[varg]{txfonts}
\usepackage{natbib}
\bibpunct{(}{)}{;}{a}{}{,}

\begin{document}

\title{Capturing transient plasma flows and jets in the solar corona}

\author{L.~P.~Chitta\inst{1}, S.~K.~Solanki\inst{1,2}, H.~Peter\inst{1}, R.~Aznar Cuadrado\inst{1}, L.~Teriaca\inst{1}, U.~Sch\"{u}hle\inst{1}, F.~Auch\`{e}re\inst{3}, D.~Berghmans\inst{4}, E.~Kraaikamp\inst{4}, S.~Gissot\inst{4} \and C.~Verbeeck\inst{4}}

\institute{Max-Planck-Institut f\"ur Sonnensystemforschung, Justus-von-Liebig-Weg 3, 37077 G\"ottingen, Germany\\
\email{chitta@mps.mpg.de}
\and
School of Space Research, Kyung Hee University, Yongin, Gyeonggi 446-701, Republic of Korea
\and
Institut d'Astrophysique Spatiale, CNRS, Univ. Paris-Sud, Universit\'{e} Paris-Saclay, B\^{a}t. 121, 91405 Orsay, France
\and
Royal Observatory of Belgium, Ringlaan -3- Av. Circulaire, 1180 Brussels, Belgium
}

   \date{Received ; accepted }

\abstract
{Intensity bursts in ultraviolet (UV) to X-ray wavelengths and plasma jets are typical signatures of magnetic reconnection and the associated impulsive heating of the solar atmospheric plasma. To gain new insights into the process, high-cadence observations are required to capture the rapid response of plasma to magnetic reconnection as well as the highly dynamic evolution of jets. Here, we report the first 2\,s cadence extreme-UV observations recorded by the 174\,\AA\ High Resolution Imager of the Extreme Ultraviolet Imager on board the Solar Orbiter mission. These observations, covering a quiet-Sun coronal region, reveal the onset signatures of magnetic reconnection as localized heating events. These localized sources then exhibit repeated plasma eruptions or jet activity. Our observations show that this spatial morphological change from localized sources to jet activity could occur rapidly on timescales of about 20\,s. The jets themselves are intermittent and are produced from the source region on timescales of about 20\,s. In the initial phases of these events, plasma jets are observed to exhibit speeds, as inferred from propagating intensity disturbances, in the range of 100\,km\,s$^{-1}$ to 150\,km\,s$^{-1}$. These jets then propagate to lengths of about 5\,Mm. We discuss examples of bidirectional and unidirectional jet activity observed to have been initiated from the initially localized bursts in the corona. The transient nature of coronal bursts and the associated plasma flows or jets along with their dynamics could provide a benchmark for magnetic reconnection models of coronal bursts and jets.}

   \titlerunning{Capturing transient plasma flows and jets in the solar corona}
   \authorrunning{L. P. Chitta et al.}

   \maketitle

\section{Introduction\label{sec:int}}

Magnetic reconnection is a universal process that is invoked to explain phenomena ranging from astrophysical jets to granular-scale explosive events on the Sun. In the solar atmosphere, signatures of magnetic reconnection are prevalent. In particular, observations of spectral line broadening and double-peaked spectral profiles support the idea of bidirectional jets during magnetic reconnection at the site of megameter-scale explosive events in the transition region at the base of the quiet-Sun corona \citep[][]{1994AdSpR..14d..13D,1997Natur.386..811I,1998ApJ...497L.109C}, as well as ultraviolet (UV) bursts in active regions \citep[][]{2014Sci...346C.315P, 2018SSRv..214..120Y}. High-cadence (1.4\,s to 5\,s) spectroscopic observations of such bursts show that the observed intensity and broadening of spectral lines fluctuate on timescales as short as 10\,s, suggesting that the plasma responds very dynamically to magnetic reconnection \citep[e.g.,][]{2015ApJ...809...82G,2020ApJ...890L...2C,2020ApJ...901..148G}. 

In chromospheric images that sample temperatures around $10^4$\,K, inverse Y-shaped anemone jets are observed in active regions. Such jets outline plasma eruptions on spatial scales of 150\,km to 300\,km, with speeds of 10\,km\,s$^{-1}$ to 20\,km\,s$^{-1}$ \citep[][]{2007Sci...318.1591S}. Signatures of similar small-scale chromospheric jets are also observed at the footpoints of active region coronal loops \citep[][]{2017ApJS..229....4C}. Small-scale plasma jets at higher speeds on the order of 100\,km\,s$^{-1}$ are observed at higher temperatures of $\sim$0.1\,MK to 1\,MK in the transition region and corona \citep[][]{2014Sci...346A.315T,2016SoPh..291.1129N,2018ApJ...868L..27P,2019ApJ...887L...8P}. Evidence of magnetic reconnection is further supported by observations that these small-scale jets erupt when patches of small-scale opposite-polarity magnetic elements in the solar photosphere interact and undergo flux cancellation \citep[e.g.][]{2017ApJS..229....4C,2018ApJ...868L..27P,2019ApJ...887L...8P}. There is also some evidence for a direct association of explosive events identified in spectroscopic observations and plasma jets observed in imaging observations \citep[][]{2013SoPh..282..453I,2019ApJ...873...79C,2021A&A...647A.159C}.

The dot-like, loop-like, and surge- or jet-like EUV brightenings in active regions, observed during the second flight of Hi-C, have provided new insights into the spatial evolution of plasma response to magnetic reconnection by revealing bi- and uni-directional plasma flows at small spatial scales of 2\arcsec--5\arcsec  \citep[][]{2019ApJ...887...56T}. However, the rapid response of high temperature plasma (above 0.1\,MK) to magnetic reconnection in terms of the spatial evolution and internal dynamics of jets are not well known due to the lack of extreme ultraviolet (EUV) coronal observations at cadences as high as 1\,s to 2\,s, which are comparable to high-cadence spectroscopic observations of the transition region \citep[e.g., at 1.4\,s,][]{2020ApJ...890L...2C}. We expect such high-cadence coronal images to reveal the transience of plasma jets as a direct response to the ephemeral nature of magnetic reconnection. We also expect such imaging observations to provide insights into the morphological evolution of jets. In this study, we report the first 2\,s high cadence observations of plasma flows or jets using data obtained by the Extreme Ultraviolet Imager \citep[EUI;][]{2020A&A...642A...8R} on board the Solar Orbiter mission \citep[][]{2020A&A...642A...1M} during its cruise phase on 2021 February 23. The reported plasma flow events are not classical or traditional coronal jets as described in the literature \citep[e.g.,][c.f. Sect.\,\ref{sec:dis}]{1994ApJ...431L..51S,2021RSPSA.47700217S}.

\section{Observations\label{sec:obs}}

\begin{figure}
\begin{center}
\includegraphics[width=0.49\textwidth]{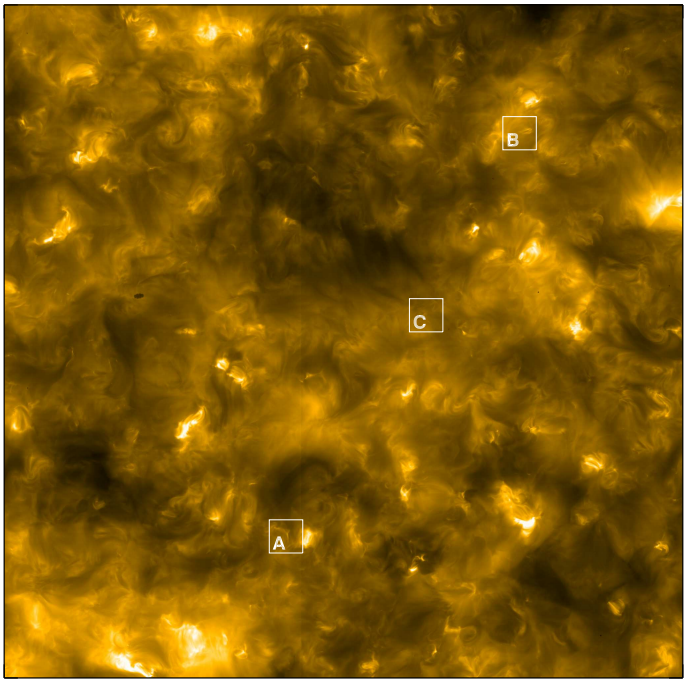}
\caption{Quiet-Sun corona observed with HRI$_{\textrm{EUV}}$ on 2021 February 23. The map is obtained by averaging emission from all the frames in the level-2 image sequence. The field of view is about 384\,Mm\,$\times$\,384\,Mm. Three boxes labelled A--C mark quiescent regions that exhibited jet activities further discussed in Figures\,\ref{fig:jet1}--\ref{fig:jet4}. Each box covers a field of view of roughly 19\,Mm\,$\times$\,19\,Mm.
\label{fig:mean}}
\end{center}
\end{figure}

\begin{figure*}
\begin{center}
\includegraphics[width=0.49\textwidth]{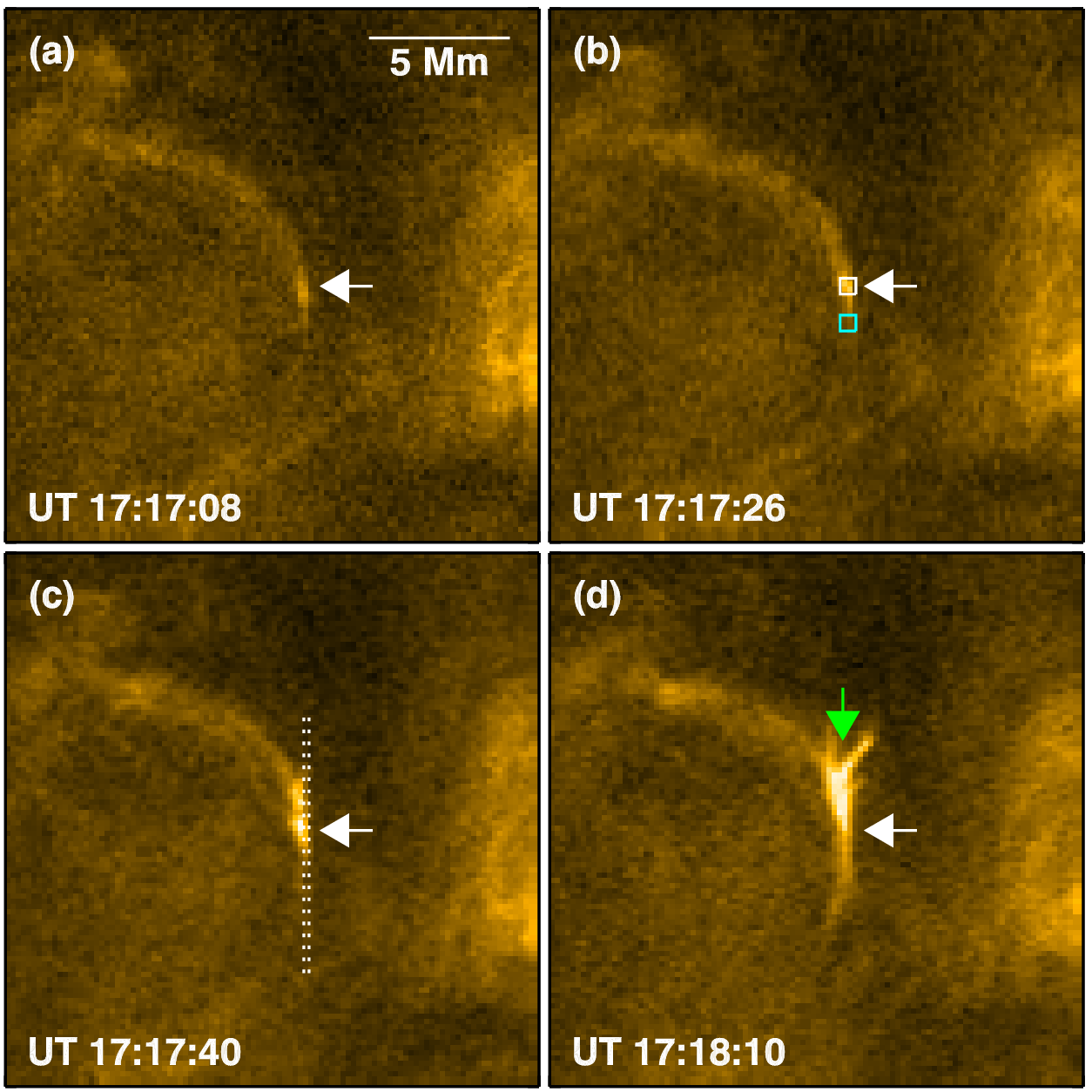}
\includegraphics[width=0.49\textwidth]{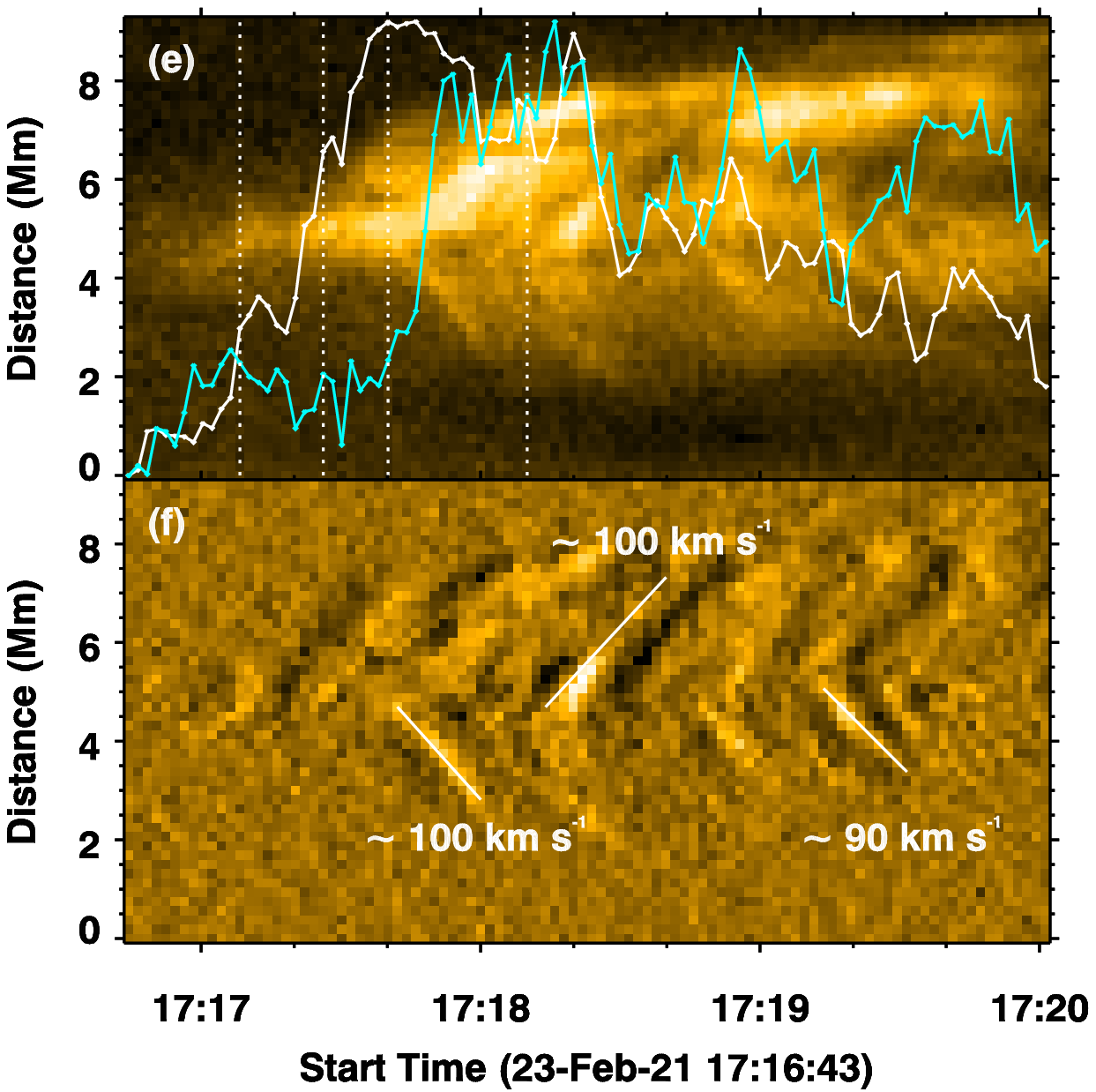}
\caption{Bidirectional jets. These transient jets are observed in the region outlined by box-A in Fig.\,\ref{fig:mean}. Panels (a)-(d), covering an area of roughly 19\,{Mm}$\times$19\,Mm, display snapshots at four instances during the jet evolution. The white arrow points to the initial brightening and its location observed prior to the jet. In panel (d), the green arrow points to a location where the jet is further bifurcated. The space-time stack plot in panel (e) shows the evolution of the brightness along the strip marked in panel (c; emission averaged across dotted vertical lines, i.e. the strip). The overlaid light curve in white is from the location of the initial brightening (emission averaged over the white box in panel b). Similarly, the cyan coloured light curve represents average emission covering a region over the jet (cyan box in panel b). The vertical dotted lines mark the time-stamps associated with images in panels (a)-(d). In panel (f) we show a smooth-subtracted image of panel (e). Fiducial (solid) lines are plotted to guide the eye. Their slopes, in units of speed (km\,s$^{-1}$), are also quoted. An online animation of the snapshots is available.\label{fig:jet1}}
\end{center}
\end{figure*}

\begin{figure*}
\begin{center}
\includegraphics[width=0.49\textwidth]{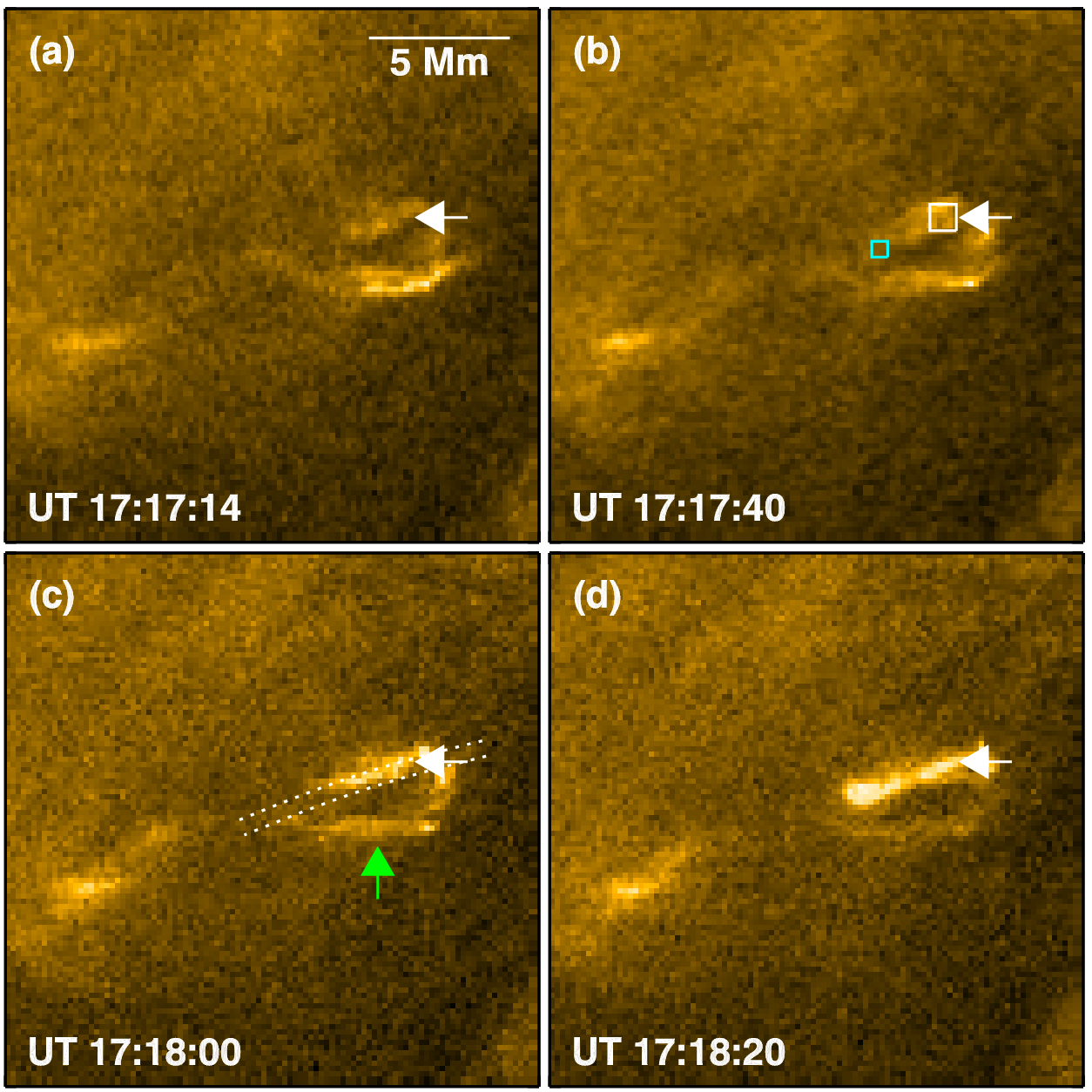}
\includegraphics[width=0.49\textwidth]{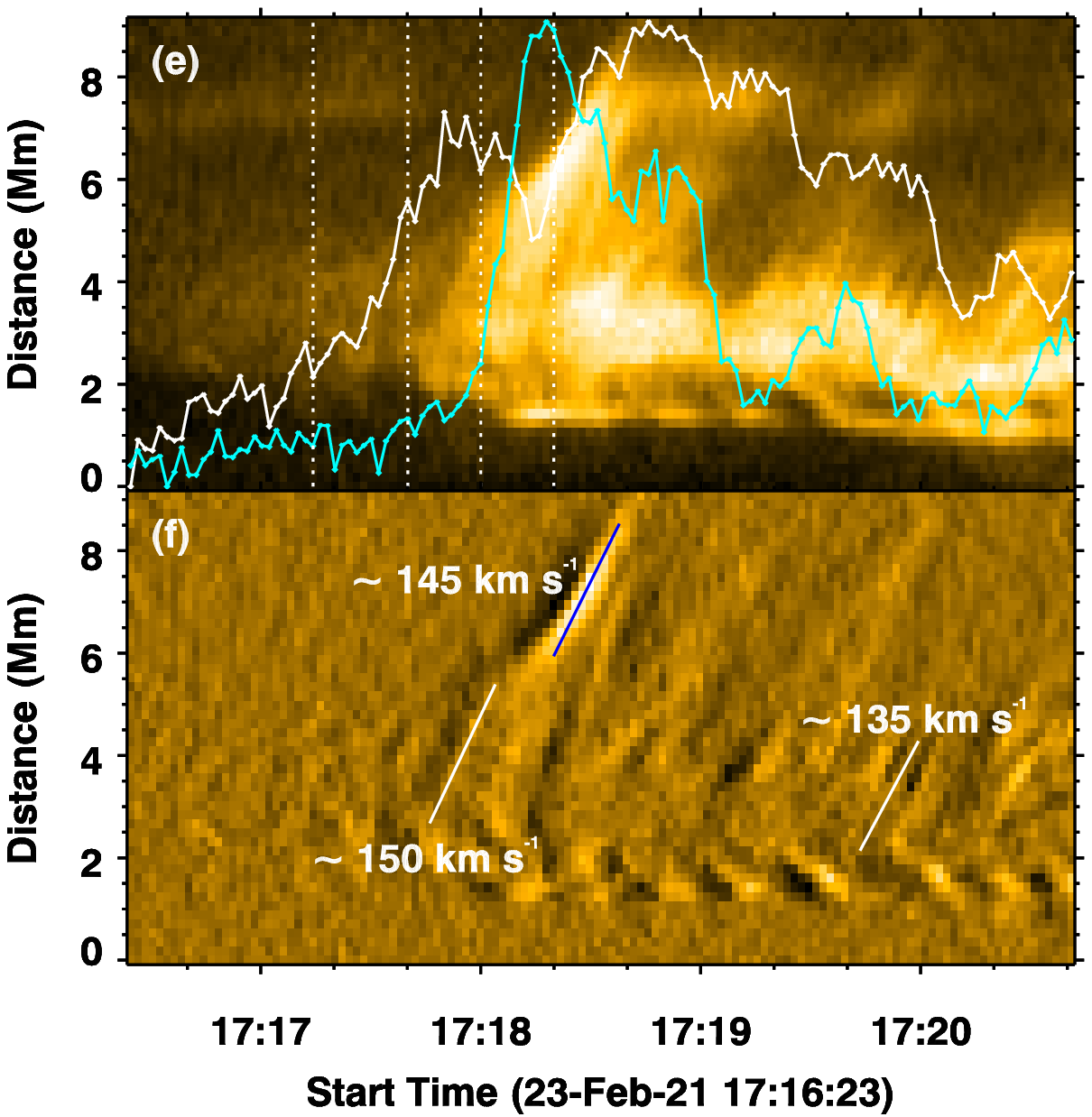}
\caption{Unidirectional jets. Same details as in Fig.\,\ref{fig:jet1}, but plotted for transient jets observed in the region outlined by box-B in Fig.\,\ref{fig:mean}. The green arrow in panel (c) points to a previously existing feature adjacent to the observed jet. An online animation of the snapshots is available.\label{fig:jet2}}
\end{center}
\end{figure*}

We employ a high-cadence image sequence recorded by the 17.4\,nm\ EUV High Resolution Imager (HRI$_{\textrm{EUV}}$) of the Extreme Ultraviolet Imager \citep[EUI;][]{2020A&A...642A...8R} on board the Solar Orbiter mission \citep[][]{2020A&A...642A...1M}, which is currently in its cruise phase. EUI/HRI$_{\textrm{EUV}}$ records the EUV emission on the central $2048\times2048$ pixel area of its CMOS sensor with an image scale of 0.492\arcsec\,pixel$^{-1}$. The observations used in this study were obtained on 2021 February 23 between UT\,17:13:25 and UT\,17:20:59 (i.e.,\ total duration of 454\,s).\footnote{\url{https://doi.org/10.24414/z2hf-b008}} At the time of observations, the sun-bound Solar Orbiter was at a distance of 0.526\,astronomical\,units and hence, the linear image scale of EUI/HRI$_{\textrm{EUV}}$ was about 188\,km\,pixel$^{-1}$. These observations were recorded at 2\,s cadence and we used level-2 calibrated data. Using a cross-correlation technique, we removed the remaining jitter in the image sequence.

EUI/HRI$_{\textrm{EUV}}$ samples emission from roughly 1\,MK plasma due to Fe\,{\sc ix} (at 17.11\,nm) and Fe\,{\sc x} (at 17.45\,nm and 17.72\,nm). In Figure\,\ref{fig:mean}, we display the mean emission map (obtained by averaging all the frames in the image sequence) covering the full field of view of HRI$_{\textrm{EUV}}$. The region covers the quiet-Sun corona that is close to the disk center as observed from the vantage point of Solar Orbiter. The Carrington longitude of Solar Orbiter was about 152\textdegree, while that of Earth was about 292\textdegree. This means that solar disk center seen from Solar Orbiter could not be seen from Earth and thus no coordinated solar observations from Earth orbit or by ground based telescopes are available.  

\section{Signatures of dynamic magnetic reconnection in the solar corona\label{sec:sig}}

The HRI$_{\textrm{EUV}}$ images revealed typical quiet-Sun coronal features such as coronal bright points on scales of 10\,Mm interspersed with patches of diffuse EUV emission (Fig.\,\ref{fig:mean}). Coronal bright points are widely studied features in the literature \citep[][]{1974ApJ...189L..93G,2019LRSP...16....2M}. The data also revealed a number of transient roundish and loop-like elongated brightenings \citep[EUV bursts or campfires; e.g.,][]{2021A&A...647A.159C,2021arXiv210403382B} that are distributed throughout the field of view (not marked in Fig.\,\ref{fig:mean}). Numerical simulations show that such transient brightenings in the corona can be triggered by magnetic (component) reconnection \citep[][]{2021arXiv210410940C}. There is also observational evidence to support magnetic reconnection as the driver of these transient coronal events based on the detection of double-peaked spectral lines (indicative of bidirectional jets from the site of reconnection) emitted by those structures at transition region temperatures of around 0.1\,MK \citep[c.f. Sect.\,\ref{sec:int}; see also][]{2021A&A...647A.159C}. 

These transient events exhibit rapid (internal) dynamics that can be seen as intensity disturbances along the brightenings in the data analyzed here. We focus on four examples of elongated brightenings detected in the 2\,s high-cadence EUI data in this study. These appear to be collimated plasma flows, which we refer to as jets in the rest of the paper. The location of these events is marked in Fig.\,\ref{fig:mean} (boxes A--C). The examples were selected based on their clear visibility, which allows us to track their evolution with minimal confusion coming from the rather diffuse and weak local fore- and background coronal emission.

\subsection{Bidirectional jet\label{sec:bid}}

Here, we analyze a bidirectional jet event that is a typical signature of magnetic reconnection. In Fig.\,\ref{fig:jet1}a-d, we show the spatial morphological evolution of a jet event. The general structure of this jet is discernible also at the center of box-A overlaid on the mean emission map in Fig.\,\ref{fig:mean}. The event started as a weak, spatially localized burst, seen as a compact bright source (indicated by an arrow in Fig.\,\ref{fig:jet1}a). The intensity of this compact source increased over a period of 20\,s (from UT\,17:17:00 to UT\,17:17:20; see white-colored light curve of the compact source in panel-e). After this phase of localized brightening, the event rapidly developed into an elongated structure (panels c-d). On one side of this elongation, the flows appears to exhibit complexity and then bifurcate further into multiple segments, diverging away from a Y-type junction (green arrow in panel d). This could reflect the internal magnetic complexity associated with this event. While the jet was visible, the compact source region exhibited further intensity fluctuations on timescales of 10\,s to 20\,s, pointing to intermittency in the magnetic reconnection. Away from the compact source, the jets are observed to exhibit intensity fluctuations on similar timescales (see cyan coloured light curve of the compact source in panel-e). The propagating nature of disturbances is apparent from the time-lag between the initial intensity rise exhibited by the compact source and that displayed by the jet.

\begin{figure*}
\begin{center}
\includegraphics[width=0.49\textwidth]{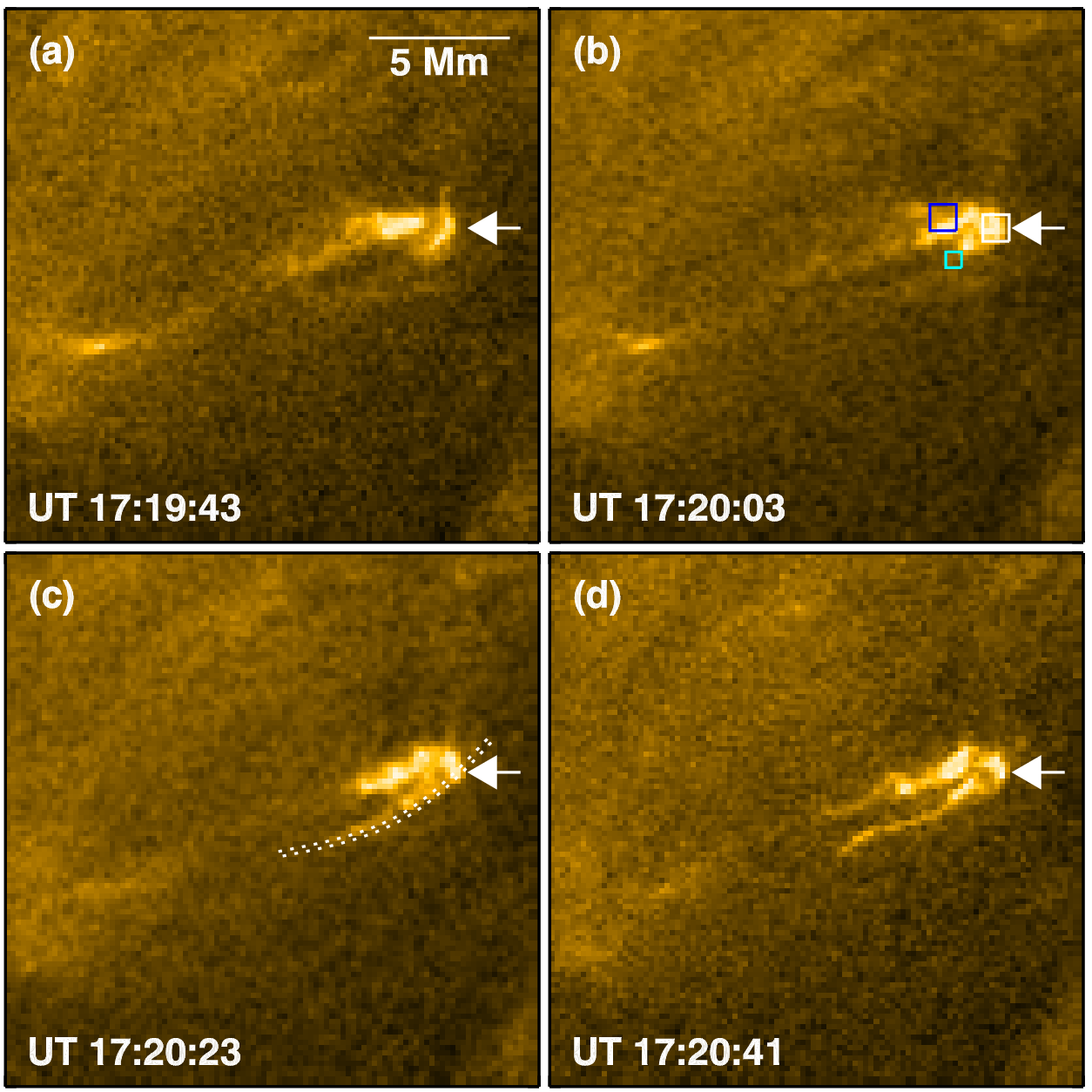}
\includegraphics[width=0.49\textwidth]{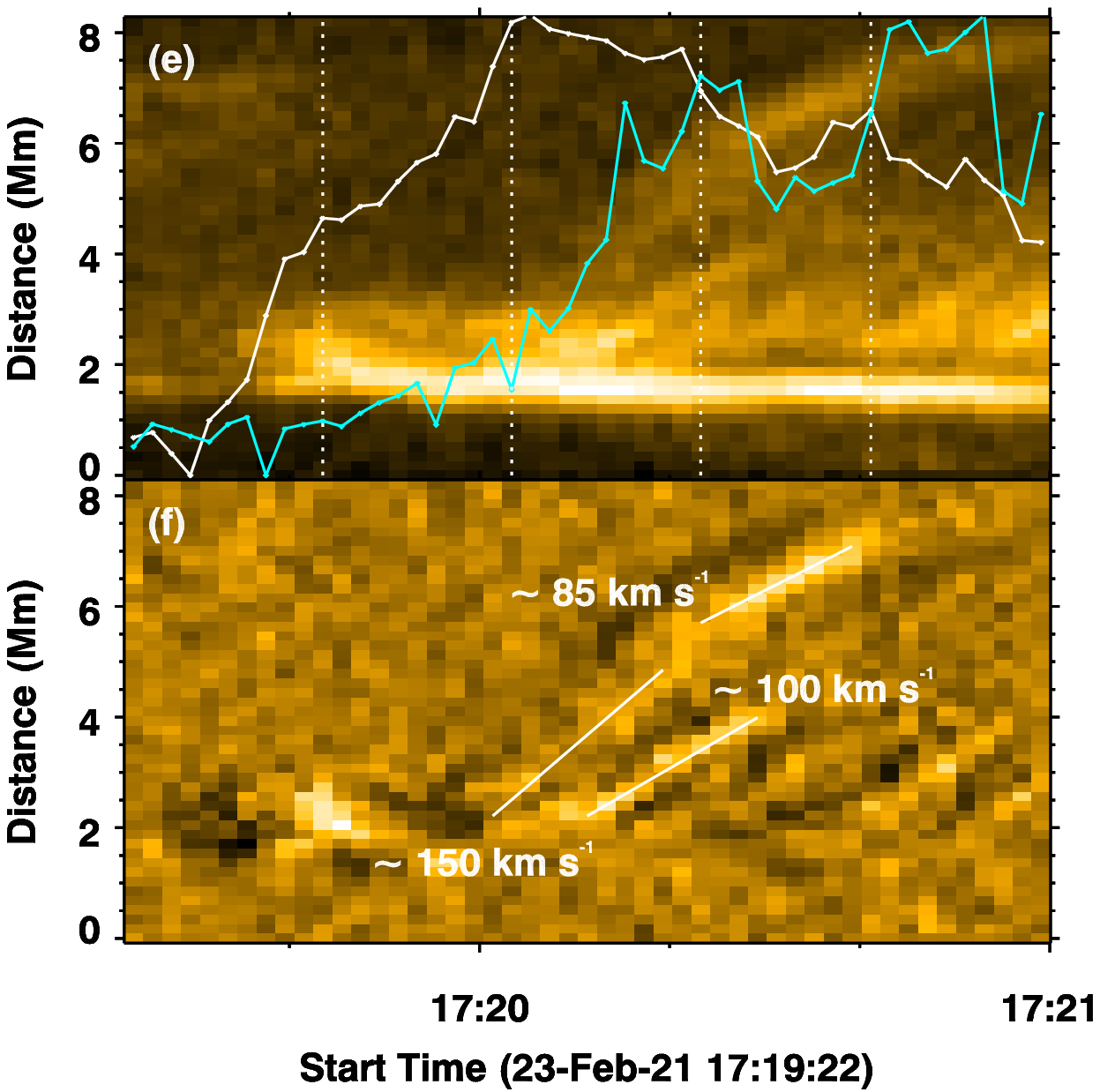}
\caption{Unidirectional jets. Same details as in Fig.\,\ref{fig:jet2}, but plotted for a second event observed in the region outlined by box-B in Fig.\,\ref{fig:mean}. For reference, the white box in Fig.\,\ref{fig:jet2}b (i.e. the source or footpoint) is overlaid as a blue box in panel-b). An online animation of the snapshots is available. \label{fig:jet3}}
\end{center}
\end{figure*}

\begin{figure*}
\begin{center}
\includegraphics[width=0.49\textwidth]{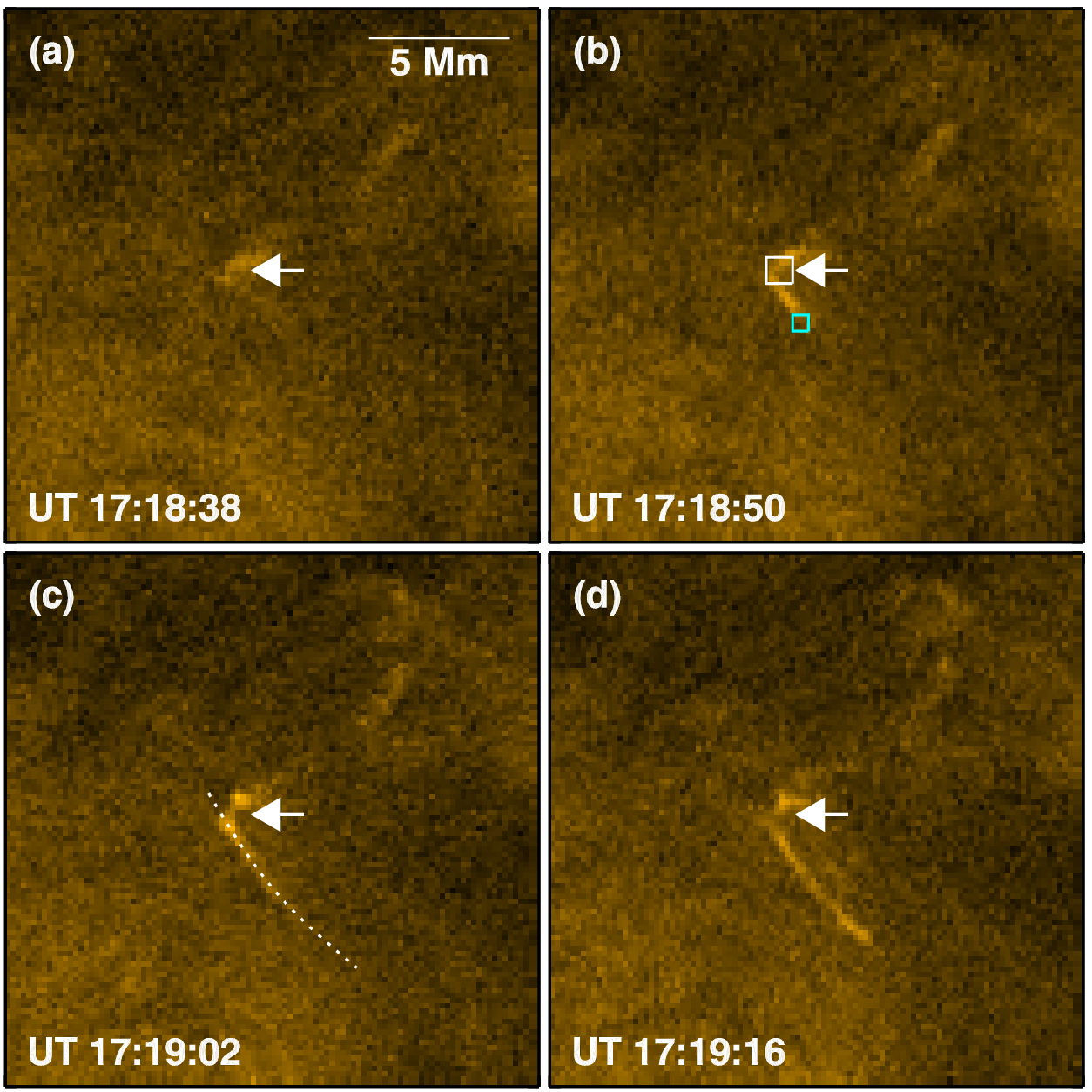}
\includegraphics[width=0.49\textwidth]{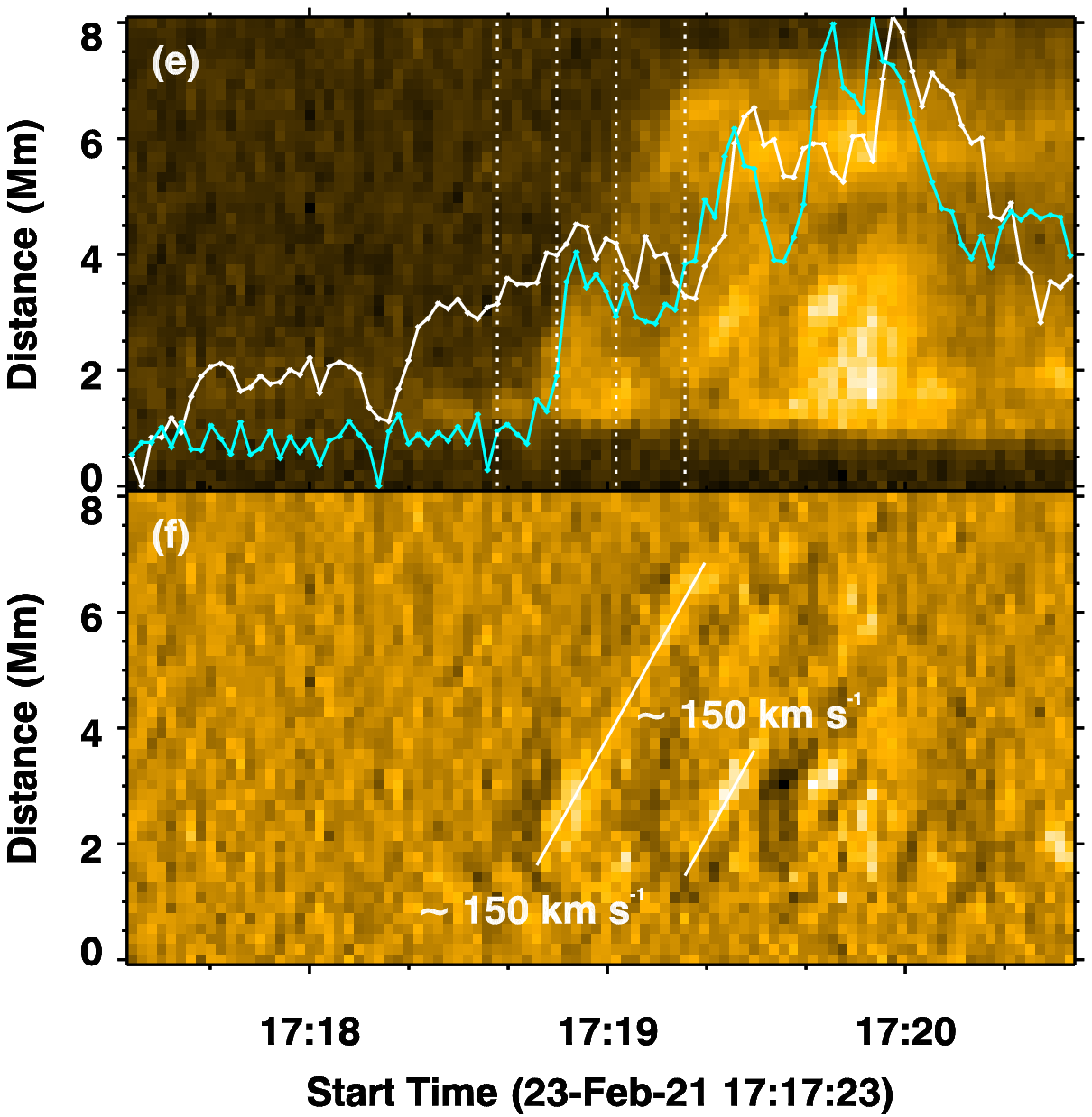}
\caption{Unidirectional jets. Same details as in Fig.\,\ref{fig:jet1}, but plotted for an event observed in the region outlined by box-C in Fig.\,\ref{fig:mean}. An online animation of the snapshots is available.\label{fig:jet4}}
\end{center}
\end{figure*}

The episodic nature of jet activity and its relation to the initial compact source are demonstrated using the space-time stack plot shown in Fig.\,\ref{fig:jet1}e, derived by plotting the intensities along a strip placed along the structure. In particular, intermittent jets are observed to originate on either side of the initial compact source (located between distances of 4\,Mm and 6\,Mm from the start of the selected strip). These disturbances apparently propagate over distances (i.e., lengths of the jets) of 2\,Mm to 4\,Mm. The intermittency of jets is further established using smooth-subtracted image of the space-time stack plot (panel-f).\footnote{To enhance the fluctuations on timescales of 20\,s observed in the light curves in panel (e), we first smoothened the space-time stack plot in the time domain using a smoothing window of 10 time steps (corresponding to a duration of 20\,s). The smoothened image is then subtracted from the original space-time map. The features remain clearly discernible over a range of smoothing windows from 5 to 25 time steps. No smooth-subtraction is done in the spatial direction. This is the case also for the smooth-subtracted images in Figs.\,\ref{fig:jet2}, \ref{fig:jet3}, and \ref{fig:jet4}. The intermittent nature of jets is evident in the associated online animations.}

Initially, these jets exhibit speeds on the order of 100\,km\,s$^{-1}$. Each episode lasts for about 20\,s, the timescale that is also observed in the intensity fluctuations of the light curve of the central bright source (panel-e). Assuming that the intensity disturbances are plasma motions, the jet speeds are comparable to the local sound speed of about 150\,km\,s$^{-1}$ in a million Kelvin hot plasma in the corona. The event  in its entirety lasted for about 200\,s.

\subsection{Unidirectional jets\label{sec:uni}}

In the following, we analyze intermittent unidirectional jets observed by HRI$_{\textrm{EUV}}$. In Fig.\,\ref{fig:jet2}a-d, we present the morphological evolution of another event that exhibits characteristics that are similar to the example in Fig.\,\ref{fig:jet1} (see also box-B in Fig.\,\ref{fig:mean}). The event in Fig.\,\ref{fig:jet2} also starts as a compact bright source (see arrow in Fig.\,\ref{fig:jet2}a). But compared to the example in Fig.\,\ref{fig:jet1}, which is more isolated, the event in Fig.\,\ref{fig:jet2} originates in a region of enhanced activity (see bright regions adjacent to the source marked by an arrow in Fig.\,\ref{fig:jet2}a). The source exhibits a rapid intensity increase for a period of roughly 20\,s (between UT\,17:17:20 to UT\,17:17:40; see light curve in panel-e). After this phase, repeated plasma eruptions are observed to originate from the source (see panels c, e-f). The propagation speeds of intensity disturbances of about 150\,km\,s$^{-1}$ are comparable to the local sound speed. This event also lasts for about 200\,s.

While the intensity disturbances propagate to relatively similar distances on either side of the central source in Fig.\,\ref{fig:jet1}f, the detected jets in Fig.\,\ref{fig:jet2}f are predominantly in one direction. On one side from the source (located about a distance of 2\,Mm from the zero position of the strip), jets have a projected length of nearly 6\,Mm. Opposite in direction to these longer jets (shown in Fig.\,\ref{fig:jet2}f), there are indications of repeated, but shorter propagating perturbations with projected lengths of only 1\,Mm. If these shorter-scale perturbations are also associated with small-scale jet activity, then the difference in the projected lengths of plasma flow in opposite directions can be understood in terms of jets propagating in closed magnetic loops, where the source region is situated closer to one of their footpoints. However, we cannot rule out other plausible explanations, such as magnetohydrodynamic waves at the footpoints of the longer jet driving these intensity perturbations on shorter length-scales of 1\,Mm.

Adjacent to the jet activity discussed in Fig.\,\ref{fig:jet2}, we observed a pre-existing loop-like plasma feature (green arrow in Fig.\,\ref{fig:jet2}c) with one of its apparent footpoints situated closer to the source region marked in Fig.\,\ref{fig:jet2}b. This loop-like structure gradually fades away by UT\,17:19:20. Its footpoint region exhibited another localized brightening (see Fig.\,\ref{fig:jet3}a-d and box-B in Fig.\,\ref{fig:mean}) that led to plasma eruptions. This burst episode exhibits brightening between UT\,17:19:40 and UT\,17:20:00, (see Fig.\,\ref{fig:jet3}e), after which point a jet is ejected at speeds of about 150\,km\,s$^{-1}$. Unlike jets seen in previous examples, that did not show signatures of speed changes as they propagate, this unidirectional jet exhibits some deceleration after UT\,17:20:20. This jet also propagates to distances of about 6\,Mm from the initial bright source. There are signatures of subsequent jet activity from this location seen as intensity ridges in the space-time stack plots (between UT\,17:20:00 and UT\,17:21:00; Fig.\,\ref{fig:jet3}e-f). The whole event evolves over a period of about 80\,s. The footpoint regions of the events shown in Figs.\,\ref{fig:jet2} and \ref{fig:jet3} appear to be close to each other (see Fig.\,\ref{fig:jet3}b). It is possible that the jet in Fig.\,\ref{fig:jet3} is causally linked to the eruption in Fig.\,\ref{fig:jet2}, which is still ongoing when the new jet starts.

An example of a jet that is fainter than the other three is presented in Fig.\,\ref{fig:jet4} (box-C in Fig.\,\ref{fig:mean}). The bright source becomes clearly distinguishable starting around UT\,17:18:20 and a jet is ejected from this region around UT\,17:18:40, which then propagates at speeds of about 150\,km\,s$^{-1}$ (Fig.\,\ref{fig:jet4}e-f, with at least two additional episodes between UT\,17:19:20 and UT\,17:20:00). Although the morphology of the jet is quite simple, its source region footpoint exhibits a dynamic evolution. The duration of this event is about 100\,s.

\section{Discussion and conclusions\label{sec:dis}}

In the initial phases, the observed transient jets propagate with speeds in the range of 100\,km\,s$^{-1}$ to 150\,km\,s$^{-1}$. Using 5\,s cadence EUI data from the 2020-May campaign, \citet[][]{2021mandal} investigated the internal dynamics of EUV brightenings \citep[termed campfires, c.f.][]{2021arXiv210403382B} and found signatures of propagating disturbances with speeds in the range of 25\,km\,s$^{-1}$ and 60\,km\,s$^{-1}$. This is a factor of two to six smaller than the speeds of jets investigated here. In this regard, the study of \citet[][]{2021mandal} complements our work in highlighting a variety of plasma dynamics in coronal structures that could be revealed, thanks to higher cadence EUI data.

The jet speeds reported in our study are on the higher end of the distribution of speeds exhibited by type-II spicules \citep[][]{2012ApJ...759...18P} and jetlet-like features from magnetic network lanes observed adjacent to an active region \citep[][]{2019ApJ...887L...8P}. The speeds, however, are generally on the lower end of transition region network jets (seen as fast propagating intensity disturbances) in coronal hole regions \citep[][]{2014Sci...346A.315T}, but comparable to transition region jets observed in the quiet-Sun \citep[][]{2016SoPh..291.1129N}. 

The main difference, however, is that the jets observed in these earlier studies appear to be collimated outflows from the Sun along (locally) open magnetic fields. In contrast, both the bi- and unidirectional jet features we presented here are apparent flows confined to closed magnetic loops in the quiet-Sun \citep[e.g.][]{2004A&A...427.1065T,2015ApJ...810...46H}. Our quiet-Sun events could be similar to the previously reported active region bi- and unidirectional surges and jet-like features \citep[][]{2019ApJ...887...56T}. 

Another related difference is the general appearance or spatial morphology of jets itself. Typical coronal jets consist of a bright base along-with a spire that extends outward from the bright jet-base into the corona \citep[][]{1994ApJ...431L..51S,1999SoPh..190..167A}. However, the collimated flows we investigated here do not show these typical characteristics of classical coronal jets. Why the studied jets differ in terms of spatial morphology in comparison with the classical coronal jets remains to be answered.

A clear imaging of bi-directional plasma jets in the corona was previously reported by \citet[][]{2013ApJ...775..132J} using EUV data from the Atmospheric Imaging Assembly \citep[AIA;][]{2012SoPh..275...17L} on board the Solar Dynamics Observatory \citep[SDO;][]{2012SoPh..275....3P}. However, the transient and repetitive nature of quiet-Sun jets on shorter timescales of 20\,s\ that we present in all examples here had not been observed before. This is because (as discussed in Appendix\,\ref{sec:app}) such short-lived transients could not be observed with routine SDO/AIA EUV observations. 

Fast-propagating disturbances in the upper chromosphere and transition region were also detected along bright elongated structures using Lyman-$\alpha$ slit-jaw images recorded by the CLASP sounding rocket experiment \citep[][]{2016ApJ...832..141K}. These CLASP Lyman-$\alpha$ propagating disturbances might be transition region counterparts to events similar to those studied here. Whether or not they are visible also in the EUI/HRI Lyman-$\alpha$ Imager requires further investigation with new observations. 

Furthermore, a common feature we observed among all four jets is that they erupt from an initial compact bright source. The intensity of the source increases for about 20\,s. The jet material is then observed as repetitive eruptions from this local bright source. The short timescales associated with these jets resemble fluctuations observed in the nonthermal broadening of transition region spectral lines (formed at around 0.1\,MK) from reconnection-driven microflares that are likely produced by internal turbulent motions  \citep[][]{2020ApJ...890L...2C}.

In three of the four jet events we investigated (i.e., Figs.\,\ref{fig:jet1}, \ref{fig:jet2}, and \ref{fig:jet4}), there is no clear signature of acceleration or deceleration of ejected material. Nevertheless, the jets fade away after traversing distances of a few Mm from the source region with almost uniform speeds. One possible explanation is that the jets are further heated or cooled down such that the resulting emission is not captured by the HRI$_{\textrm{EUV}}$ 174\,\AA\ filter. In one case (Fig.\,\ref{fig:jet3}), however, the initial jet is observed to slow down. Whether this deceleration reflects the true kinematics of the jet plasma or whether it is an apparent effect (i.e., produced by the loop geometry that curves perpendicular to the plane of sky, as in a downward curved loop) requires further investigation with more such events, coupled with magnetic field extrapolations.

Because these jets are characterized by their intensity enhancement with respect to the local background, it is either the segments of these jets or the whole structures themselves that could be classified as quiet-Sun EUV bursts \citep[][]{2021A&A...647A.159C} or campfires \citep[][]{2021arXiv210403382B}, which are likely akin to small-scale bursts detected in active regions \citep[][]{2019ApJ...887...56T}. Our spatio-temporal analysis provides further insights into the dynamics and transient nature of these small-scale events.

Magnetic reconnection models of solar bursts predict that the current sheets break into copious magnetic islands or plasmoids that are then ejected outward. These plasmoids then hit the ambient medium and thermalize to produce repeated bursts  of observable plasma emission  \citep[][]{2019A&A...628A...8P,2020ApJ...901..148G}.  It is possible that the jets we observed are related to the dynamics of such plasmoids along reconnection current sheets in coronal bursts. The observed speeds of jets, their transient nature (short lifetimes) and their origination from initially localized bursts could be used to test magnetic reconnection models of coronal bursts and jets.  

The data we used here are among the highest cadence EUV imaging observations of the solar corona ever recorded. These novel 2\,s\ cadence data revealed the repetitive and transient nature of coronal jets on timescales of about 20\,s. There are hints of similar plasma flows (with repetition) also in coronal bright points distributed throughout the field of view. However, most of these regions appear over-exposed in this experimental observational set, limiting the extraction of local flow properties. Similar high-cadence EUV observations (with better exposure settings to capture bright regions) during the perihelion of Solar Orbiter (around 0.3\,AU) with roughly two times better spatial resolution than the data used here will further reveal and establish the details of the internal dynamics and the transient nature of these jets and their relation to plasmoids. Combining such observations with photospheric magnetic field information at a similar spatial resolution that is expected from the Polarimetric and Helioseismic Imager on Solar Orbiter \citep[][]{2020A&A...642A..11S} will shed light on the magnetic drivers of these transient jets.

\begin{acknowledgements}
We thank the anonymous referee for helpful comments on the manuscript. The authors thank Koen Stegen (Royal Observatory of Belgium) for his crucial role in EUI operations. L.P.C. thanks Sudip Mandal (MPS) for helpful discussions. This project has received funding from the European Research Council (ERC) under the European Union’s Horizon 2020 research and innovation programme (grant agreement No 695075). Solar Orbiter is a mission of international cooperation between ESA and NASA, operated by ESA. The EUI instrument was built by CSL, IAS, MPS, MSSL/UCL, PMOD/WRC, ROB, LCF/IO with funding from the Belgian Federal Science Policy Office (BELSPO/PRODEX PEA 4000112292); the Centre National d’Etudes Spatiales (CNES); the UK Space Agency (UKSA); the Bundesministerium f\"{u}r Wirtschaft und Energie (BMWi) through the Deutsches Zentrum f\"{u}r Luft- und Raumfahrt (DLR); and the Swiss Space Office (SSO).
\end{acknowledgements}

\begin{appendix}
\section{SDO/AIA type observations of jets\label{sec:app}}
\begin{figure*}
\begin{center}
\includegraphics[width=0.49\textwidth]{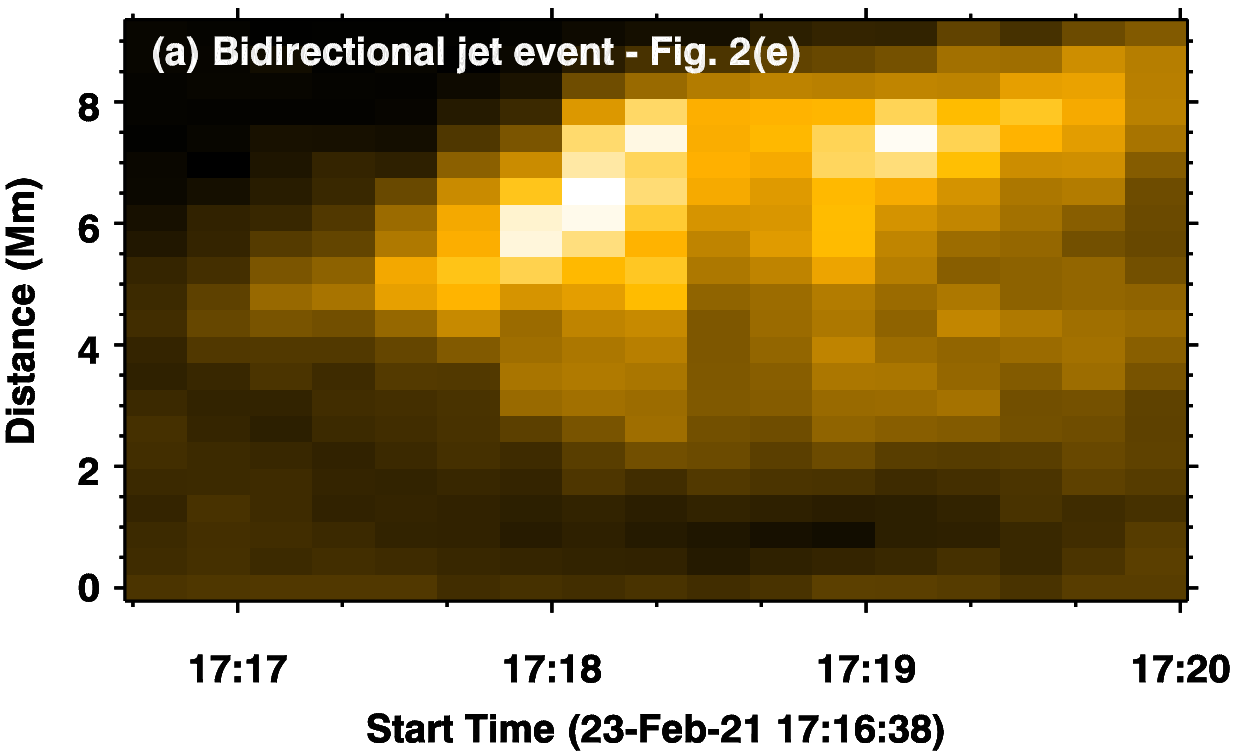}
\includegraphics[width=0.49\textwidth]{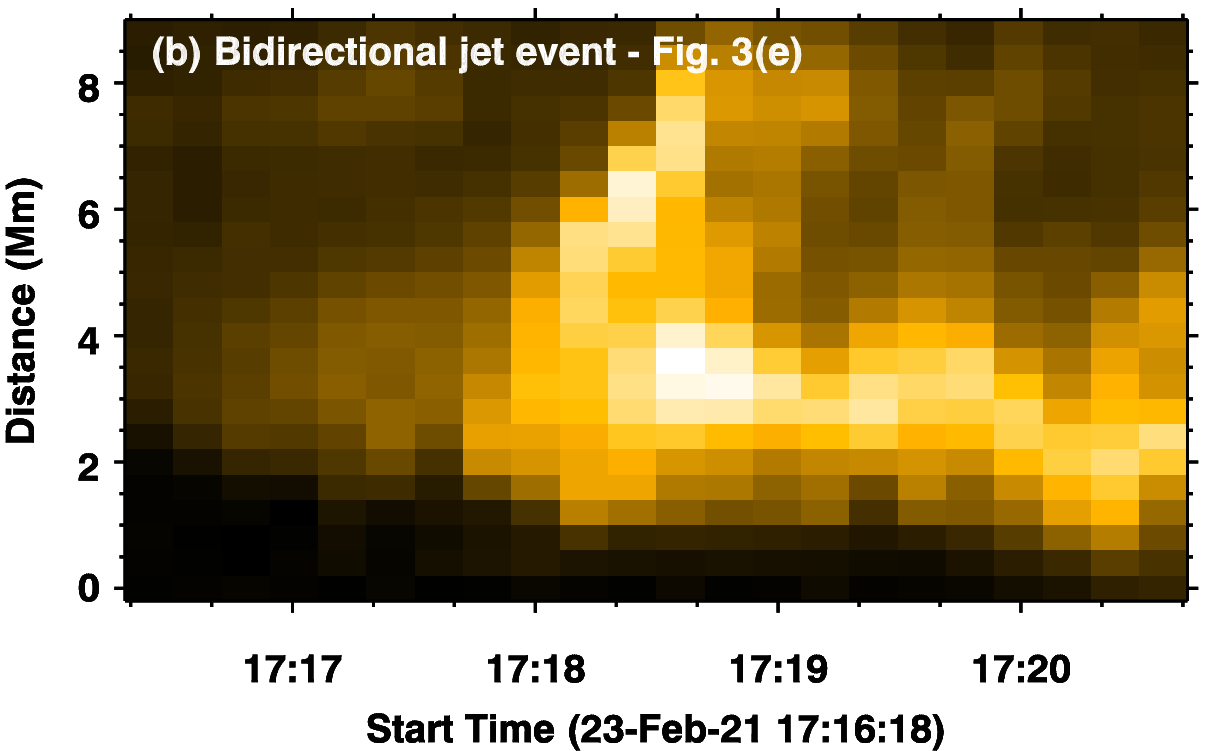}
\includegraphics[width=0.49\textwidth]{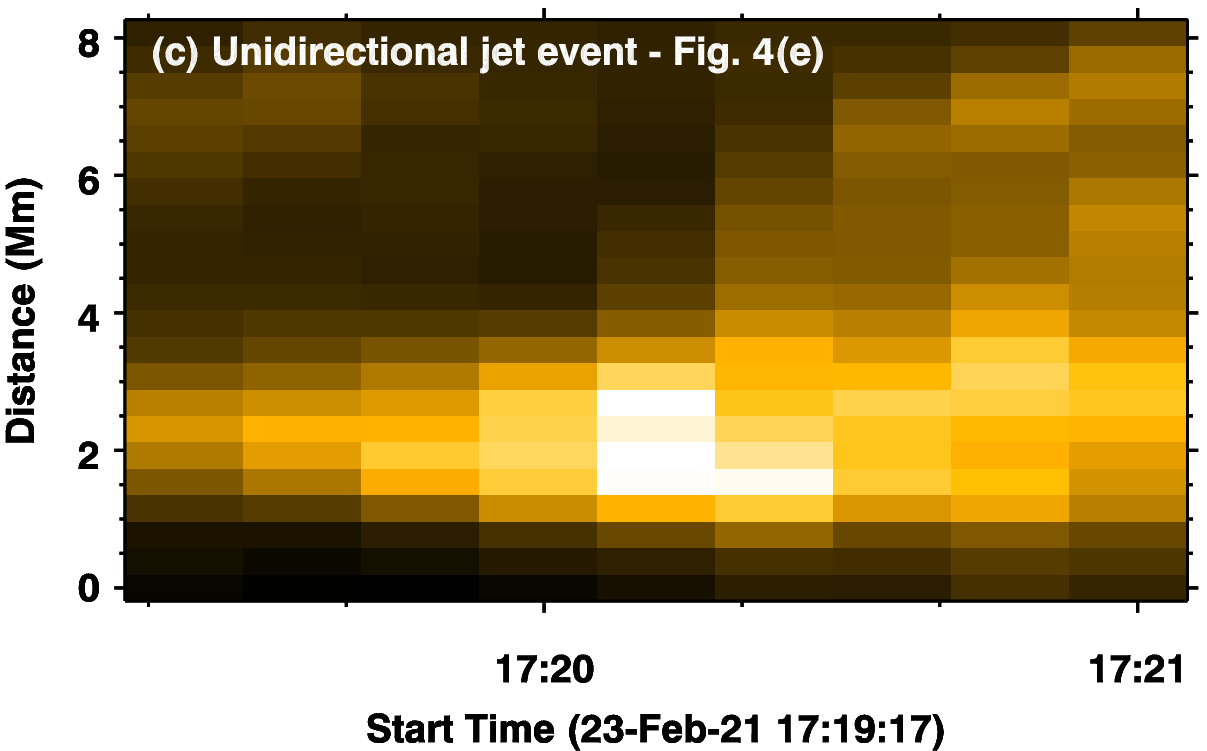}
\includegraphics[width=0.49\textwidth]{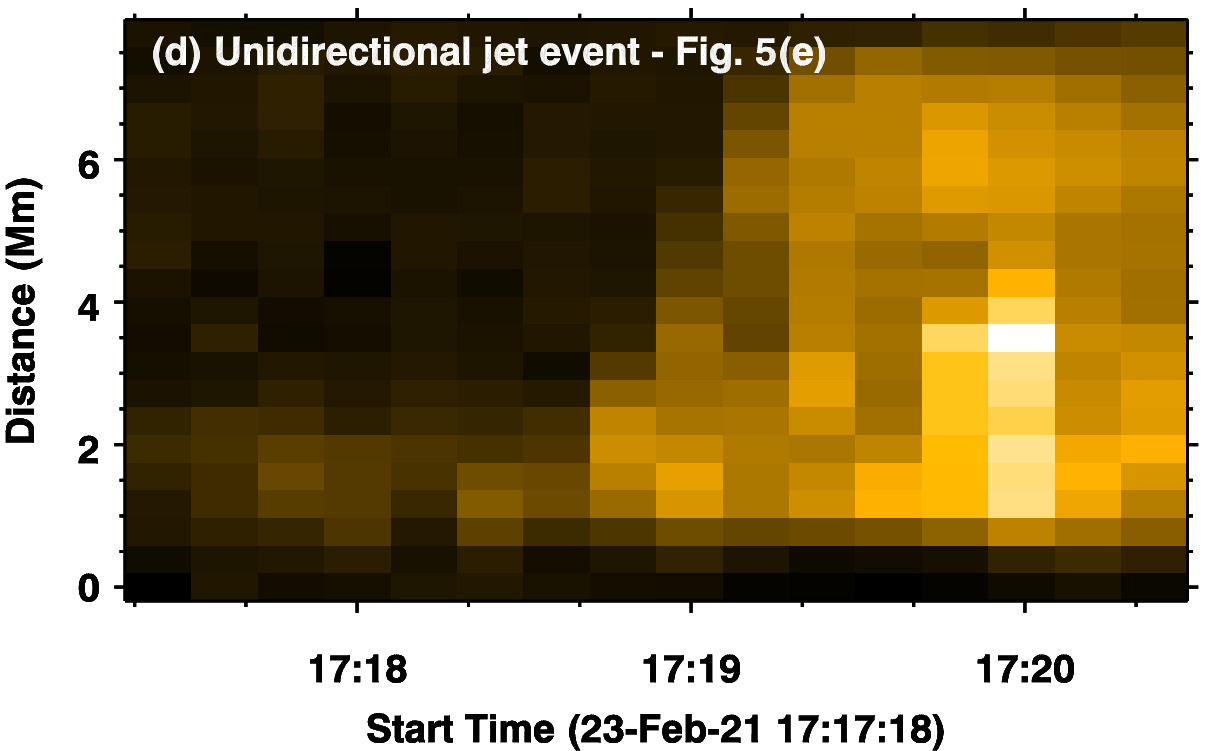}
\caption{Jet features as SDO/AIA would observe them. Panels (a) through (d) show space-time stack plots of jet features that correspond to panel (e) in Fig.\,\ref{fig:jet1} through Fig.\,\ref{fig:jet4}, as if those jets were observed by EUV filters on SDO/AIA that sample plasma emission from similar temperatures as HRI$_{\textrm{EUV}}$ would (e.g.,\ AIA 17.1\,nm or 19.3\,nm filters; but with six times lower cadence at roughly a factor of two lower spatial resolution).\label{fig:jetaia}}
\end{center}
\end{figure*}

The four cases we discussed in the main text have jet propagation lengths of about 5\,Mm and total event duration exceeding 60\,s. These characteristics should imply that they (or similar events) are observable with the Atmospheric Imaging Assembly \citep[AIA;][]{2012SoPh..275...17L} on board the Solar Dynamics Observatory \citep[SDO;][]{2012SoPh..275....3P}. The EUV channels of SDO/AIA record images every 12\,s and have a spatial resolution of about 1.2\arcsec. Therefore, it may be asked why such events were not reported based on AIA observations. At the time the observations were recorded, Solar Orbiter was close to the far-side of the Sun (c.f. Sect.\,\ref{sec:obs}), and therefore no direct SDO observations were available. Therefore, to answer the question, we used the EUI data of the observed jets themselves and retrieved their space-time stack plots for parameters typical of observations by SDO/AIA channels, which record image sequences of the solar corona at a six times lower cadence and a roughly two times lower spatial resolution compared to the HRI$_{\textrm{EUV}}$ images presented in our study. To simulate the six times lower cadence of AIA, we simply picked every sixth EUI snapshot. The next step is to spatially degrade the EUI data to closely match the spatial resolution of AIA. To this end, we extracted the central $9\times9$ pixels patch covering the core of the point spread function (PSF) of the AIA 171\,\AA\ filter. This patch was then rescaled and the resulting 2D profile was normalized such that the target PSF or kernel has an image scale similar to that of EUI and area under the PSF equals unity (to preserve flux). We first convolved the EUI images with the newly created kernel and then rebinned the data such that the image scale is 0.6\arcsec, matching the AIA plate scale. Using these degraded data, we constructed the space-time maps of the jet features. The results are displayed in Fig.\,\ref{fig:jetaia}. This experiment reveals that while a general intensity propagation along the structures is observed by SDO/AIA, it would miss most of the temporal substructure and to a large extent the rapid evolution of jets we observed with HRI$_{\textrm{EUV}}$. For instance, the clear transient bidirectional jet features that we discussed in Fig.\,\ref{fig:jet1} are not apparent in Fig.\,\ref{fig:jetaia}a. Moreover, the repetition of transient jet activity (on timescales of 20\,s) that we observed in all four cases is also clearly missing in the SDO/AIA type  observations, which means missing signatures of dynamic magnetic reconnection. Thus, our study emphasizes the necessity of high-cadence, high-resolution observations to capture the highly dynamic signatures of magnetic reconnection in the solar corona.

\end{appendix}

\end{document}